\documentclass{article}
\usepackage{iclr2026_conference}
\iclrfinalcopy
\usepackage[utf8]{inputenc}
\usepackage[T1]{fontenc}
\usepackage{hyperref}
\usepackage{url}
\usepackage{booktabs}
\usepackage{amsfonts}
\usepackage{amsmath}
\usepackage{amsthm}
\theoremstyle{remark}
\newtheorem{remark}{Remark}
\usepackage{nicefrac}
\usepackage{microtype}
\usepackage{xcolor}
\usepackage{graphicx}
\usepackage{enumitem}
\usepackage{multirow}
\usepackage{tabularx}
\usepackage{subcaption}
\usepackage{natbib}
\usepackage{fancyhdr}

\usepackage{tikz}

\newcommand{\pixelclaude}[1][0.035]{%
\begin{tikzpicture}[scale=#1, baseline=-2pt]
\fill[color={rgb,255:red,139;green,94;blue,60}] (2,12) rectangle (3,13);
\fill[color={rgb,255:red,139;green,94;blue,60}] (3,12) rectangle (4,13);
\fill[color={rgb,255:red,139;green,94;blue,60}] (4,12) rectangle (5,13);
\fill[color={rgb,255:red,139;green,94;blue,60}] (5,12) rectangle (6,13);
\fill[color={rgb,255:red,139;green,94;blue,60}] (1,11) rectangle (2,12);
\fill[color={rgb,255:red,139;green,94;blue,60}] (2,11) rectangle (3,12);
\fill[color={rgb,255:red,139;green,94;blue,60}] (3,11) rectangle (4,12);
\fill[color={rgb,255:red,139;green,94;blue,60}] (4,11) rectangle (5,12);
\fill[color={rgb,255:red,139;green,94;blue,60}] (5,11) rectangle (6,12);
\fill[color={rgb,255:red,139;green,94;blue,60}] (6,11) rectangle (7,12);
\fill[color={rgb,255:red,255;green,218;blue,185}] (1,10) rectangle (7,11);
\fill[color={rgb,255:red,255;green,218;blue,185}] (1,9) rectangle (7,10);
\fill[black] (2,9) rectangle (3,10);
\fill[black] (5,9) rectangle (6,10);
\fill[color={rgb,255:red,74;green,158;blue,218}] (1,7) rectangle (7,9);
\fill[color={rgb,255:red,74;green,158;blue,218}] (0,6) rectangle (8,7);
\fill[color={rgb,255:red,74;green,158;blue,218}] (1,5) rectangle (7,6);
\fill[color={rgb,255:red,255;green,218;blue,185}] (0,7) rectangle (1,9);
\fill[color={rgb,255:red,255;green,218;blue,185}] (7,7) rectangle (8,9);
\fill[color={rgb,255:red,70;green,70;blue,70}] (1,3) rectangle (3,5);
\fill[color={rgb,255:red,70;green,70;blue,70}] (5,3) rectangle (7,5);
\fill[color={rgb,255:red,50;green,50;blue,50}] (1,2) rectangle (3,3);
\fill[color={rgb,255:red,50;green,50;blue,50}] (5,2) rectangle (7,3);
\end{tikzpicture}%
}

\newcommand{\pixelgpt}[1][0.035]{%
\begin{tikzpicture}[scale=#1, baseline=-2pt]
\fill[color={rgb,255:red,44;green,44;blue,44}] (2,13) rectangle (6,14);
\fill[color={rgb,255:red,44;green,44;blue,44}] (1,12) rectangle (7,13);
\fill[color={rgb,255:red,255;green,218;blue,185}] (1,11) rectangle (7,12);
\fill[color={rgb,255:red,255;green,218;blue,185}] (1,10) rectangle (7,11);
\fill[color={rgb,255:red,85;green,85;blue,85}] (1.5,10) rectangle (3.5,11);  
\fill[color={rgb,255:red,85;green,85;blue,85}] (4.5,10) rectangle (6.5,11);  
\fill[color={rgb,255:red,85;green,85;blue,85}] (3.5,10.3) rectangle (4.5,10.7);  
\fill[color={rgb,255:red,255;green,218;blue,185}] (1,9) rectangle (7,10);
\fill[color={rgb,255:red,212;green,149;blue,106}] (3,9) rectangle (5,10);  
\fill[color={rgb,255:red,116;green,170;blue,99}] (1,7) rectangle (7,9);
\fill[color={rgb,255:red,116;green,170;blue,99}] (0,6) rectangle (8,7);
\fill[color={rgb,255:red,116;green,170;blue,99}] (1,5) rectangle (7,6);
\fill[color={rgb,255:red,255;green,218;blue,185}] (0,7) rectangle (1,9);
\fill[color={rgb,255:red,255;green,218;blue,185}] (7,7) rectangle (8,9);
\fill[color={rgb,255:red,68;green,68;blue,68}] (1.5,3) rectangle (3.5,5);
\fill[color={rgb,255:red,68;green,68;blue,68}] (4.5,3) rectangle (6.5,5);
\fill[color={rgb,255:red,50;green,50;blue,50}] (1.5,2) rectangle (3.5,3);
\fill[color={rgb,255:red,50;green,50;blue,50}] (4.5,2) rectangle (6.5,3);
\end{tikzpicture}%
}

\newcommand{\aris}{\textsc{Aris}}
\newcommand{\skillmd}{\texttt{SKILL.md}}
\newcommand{\ariscode}{\textsc{Aris-Code}}

\title{%
\pixelclaude[0.06]\quad
ARIS: Autonomous Research via\\Adversarial Multi-Agent Collaboration%
\quad\pixelgpt[0.06]%
}

\author{
\textbf{Ruofeng Yang}$^{1\dagger}$, \textbf{Yongcan Li}$^{1}$, \textbf{Shuai Li}$^{1,2*}$ \\[3pt]
$^1$Shanghai Jiao Tong University \quad $^2$Shanghai Innovation Institute \\[2pt]
\texttt{\{wanshuiyin, joseph\_y, shuaili8\}@sjtu.edu.cn} \\[3pt]
$^\dagger$Project Leader \quad $^*$Corresponding Author \\[2pt]
{\small Project page: \url{https://github.com/wanshuiyin/Auto-claude-code-research-in-sleep}}
}

\begin{document}

\maketitle
\thispagestyle{fancy}

\vspace{-0.8cm}
\begin{center}
    \includegraphics[width=0.95\textwidth]{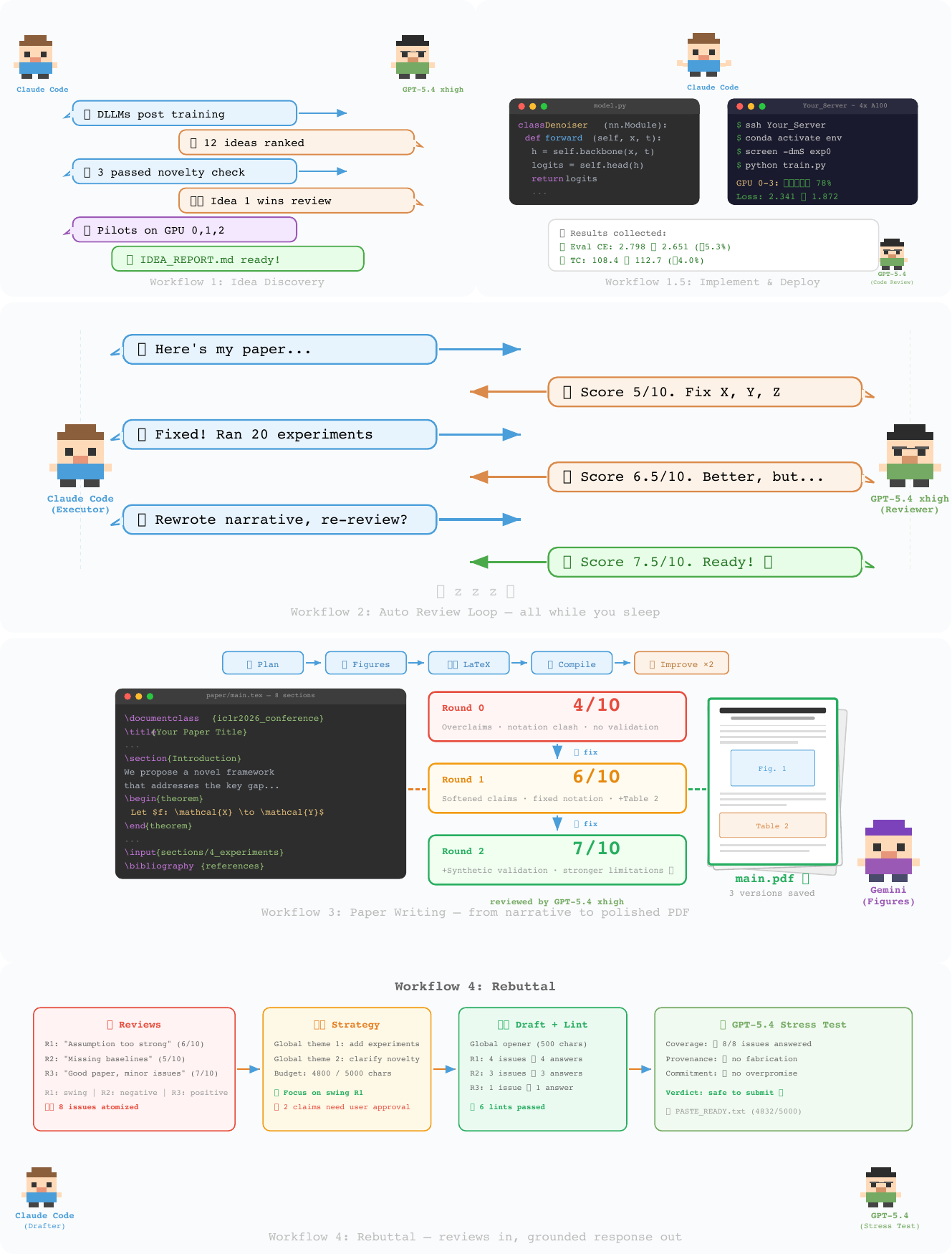}
\end{center}

\begin{abstract}
This report describes \aris{}(\textbf{A}utonomous \textbf{R}esearch via Adversarial Multi-Agent Collaboration), an open-source research harness for autonomous ML research, including its architecture, assurance mechanisms, and early deployment experience.The performance of agent systems built on large language models depends on both model weights and the \emph{harness} around them, which is the system logic that governs what information to store, retrieve, and present to the model.
For long-horizon research workflows, the central failure mode is not visible breakdown but \emph{plausible unsupported success}: a long-running agent can produce claims whose evidential support is incomplete, misreported, or silently inherited from the executor's framing.
Therefore, we present \aris{} as a research harness that coordinates machine-learning research workflows through cross-model adversarial collaboration as a default configuration: an executor model drives forward progress while a reviewer from a different model family is recommended to critique intermediate artifacts and request revisions.
\aris{} has three architectural layers.
The \emph{execution layer} provides more than 65 reusable Markdown-defined skills, model integrations via MCP, a persistent research wiki for iterative reuse of prior findings, and deterministic figure generation.
The \emph{orchestration layer} coordinates five end-to-end workflows with adjustable effort settings and configurable routing to reviewer models.
The \emph{assurance layer} includes a three-stage process for checking whether experimental claims are supported by evidence—integrity verification, result-to-claim mapping, and claim auditing that cross-checks manuscript statements against the claim ledger and raw evidence—as well as a five-pass scientific-editing pipeline, mathematical-proof checks, and visual inspection of the rendered PDF.
A prototype self-improvement loop records research traces and proposes harness improvements that are adopted only after reviewer approval.

\end{abstract}

\section{Introduction}
\label{sec:intro}

Recent work on harness engineering~\citep{lee2026metaharness} suggests that the performance of LLM systems can depend heavily on the \emph{harness}---the surrounding system logic that governs storage, retrieval, and presentation---as well as on model weights.
Machine-learning research poses an unusually complex harness-engineering problem: the workflow spans literature review and hypothesis generation through experimentation, internal critique, manuscript preparation, and responses to external feedback.
This research harness is still assembled manually in many settings: researchers coordinate compute, references, manuscript tooling, and feedback workflows across separate systems~\citep{lu2024aiscientist,schmidgall2025agentrxiv}.

Several autonomous research agents now target specific parts of this workflow.
The AI Scientist~\citep{lu2024aiscientist} and AI Scientist v2~\citep{yamada2025aiscientistv2} automate a pipeline from idea generation to manuscript drafting.
Agent Laboratory~\citep{schmidgall2025agentrxiv}
adds human-in-the-loop checkpoints to the workflow.
These systems exhibit three recurring limitations that motivate our design:
(1)~many rely on the same or closely related model family for both execution and review---a same-model self-refinement pattern in the spirit of~\citet{madaan2023selfrefine,shinn2023reflexion}---which can leave correlated errors uncaught when generator and validator share inductive biases (an effect that motivates work on heterogeneous multi-agent debate~\citealp{du2024multiagentdebate,liang2024divergent});
(2)~workflows are tightly coupled end-to-end, making it difficult to replace individual stages or resume from saved intermediate states;
(3)~few provide explicit, system-level checks on experimental integrity and manuscript quality.

As current agents become more capable of carrying out long-horizon tasks, it is possible to conduct fully autonomous research from an intuition or a basic idea. However, when using a single agent to conduct a long-term hard task, it may exhibit laziness, hallucinations, or deceptive behavior.
The central risk for an autonomous research harness is not only outright failure, but \emph{plausible unsupported success}: results may be real yet misreported, claims may outrun the evidence that licenses them, and downstream readers may silently inherit the executor's framing. Hence, we propose the following stringent assumption:
\begin{center}
    \textit{Any long-term task performed by a single agent is unreliable.}\\
        \textit{We need to divide the total workflow into sub-workflows and cross-family models to review the output at each step independently.}
\end{center}
This assumption may understate the capabilities of current agents, but the trade-off favors strictness in a high-rigor field like research: an adversarial reviewer offers a clear quality gain even though adversarial review introduces a harder optimization problem for the executor. Think of it as adversarial vs.\ stochastic bandits---a single model self-reviewing is the stochastic case (predictable reward noise), while cross-model review is adversarial (the reviewer actively probes weaknesses the executor did not anticipate), and adversarial bandits are fundamentally harder to game. Two agents (executor and reviewer) are also the minimum needed to break self-play blind spots, and two-player games converge to a Nash equilibrium far more efficiently than $n$-player ones.

This stringent assumption decomposes operationally into three bottlenecks.
First, \emph{persistent research state}~(i) is required because stepwise review is meaningless if the system cannot preserve the artifacts, decisions, evidence, and claims that connect one sub-workflow to the next.
Second, \emph{modular execution}~(ii) is required because a long research trajectory must be divided into replaceable stages rather than hidden inside a single opaque agent trajectory.
Third, \emph{independent assurance}~(iii) is required because the reviewer must not merely continue the executor's reasoning, but examine the produced artifact from a sufficiently different model family, context policy, or audit role.
These are not separate desiderata added after the fact; they are the system-level consequences of treating single-agent long-horizon research as unreliable by default.

\aris{} responds by treating assurance as a first-class workflow layer rather than a single review pass, separating artifact production from evidence checking, claim mapping, and manuscript review. Concretely, reusable Markdown-defined skills are coordinated under a default cross-family executor/reviewer pairing, with explicit assurance checks at key experimental and manuscript stages. We default to cross-family pairings because prior work suggests that mixed-model agent configurations can produce less correlated and more varied critiques~\citep{du2024multiagentdebate,liang2024divergent}; we adopt this as a recommended configuration rather than a hard system constraint.

We describe three aspects of \aris{}:
\begin{enumerate}[nosep,leftmargin=*]
    \item An \textbf{assurance stack} that uses separate executor and reviewer models, including a three-stage process for checking whether claims are supported by evidence (integrity verification, result-to-claim mapping, claim auditing against the claim ledger and raw evidence), a five-pass scientific-editing pipeline, mathematical-proof checks, and visual PDF inspection (\S\ref{sec:assurance}).
    \item A \textbf{modular system architecture} organized into three layers---execution, orchestration, and assurance---with more than 65 reusable skills, a persistent research wiki for iterative reuse of prior findings, deterministic figure generation, adjustable effort levels, configurable reviewer routing, and a prototype self-improvement loop (\S\ref{sec:overview}--\S\ref{sec:metaopt}).
    \item \textbf{Early deployment experience} across three tested executor platforms with adaptation guides for three additional platforms, including community usage reports and an analysis of current limitations (\S\ref{sec:evidence}).
\end{enumerate}
\begin{remark}[Discussion of Human in the Loop]
    Though \aris{} is an auto-research system, we still note that human-in-the-loop can significantly improve the generation quality of final papers and can help users to gain more knowledge of writing papers, which is essential for cultivating one's research taste. 
\end{remark}

\section{System Overview}
\label{sec:overview}

\begin{figure*}[t]
    \centering
    \includegraphics[width=0.95\textwidth]{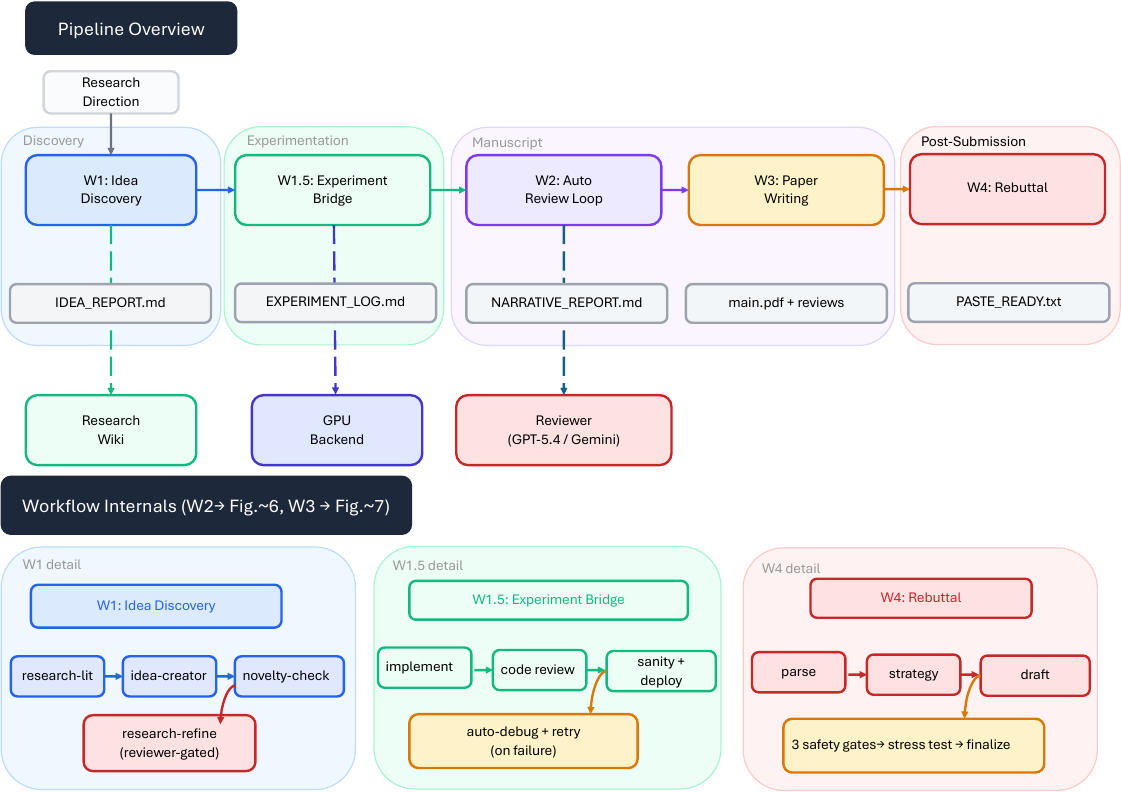}
    \caption{\aris{} workflow library. \textbf{Top}: end-to-end composition of the five workflows and their artifact contracts, grouped into four research phases (Discovery, Experimentation, Manuscript, Post-Submission); dashed links denote reviewer feedback, GPU-triggered evidence collection, and wiki memory. \textbf{Bottom}: compressed internal structure for the workflows not otherwise expanded in the main text---W1 idea discovery (with reviewer-gated refinement), W1.5 experiment bridge (with code review and auto-debug fallback), and W4 rebuttal (with safety gates and stress test). W2 auto-review and W3 paper writing internals are detailed separately in Figures~\ref{fig:wf2} and~\ref{fig:wf3}.}
    \label{fig:pipeline}
\end{figure*}

\begin{figure*}[t]
    \centering
    \includegraphics[width=0.95\textwidth]{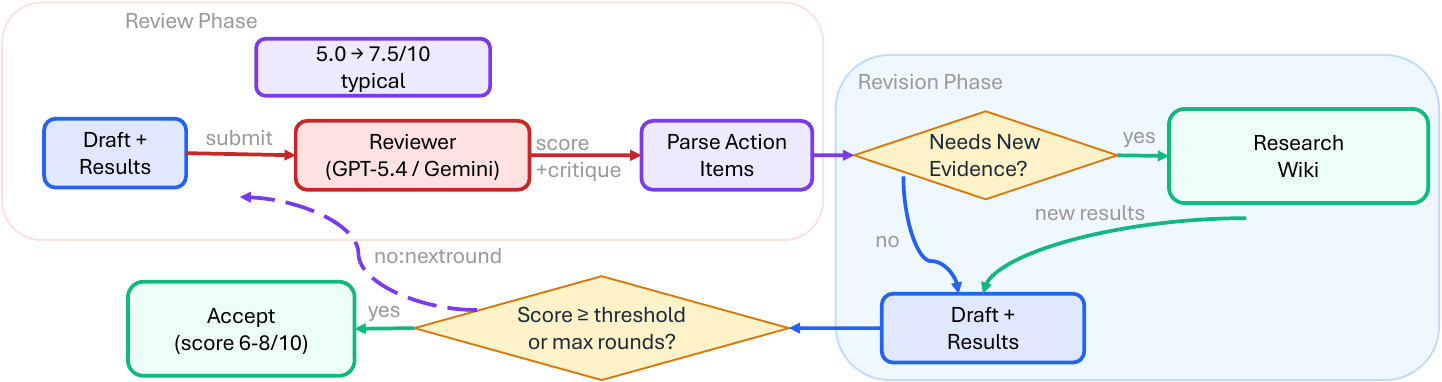}
    \caption{Workflow~2: Auto Review Loop. Each round submits the draft to a cross-model reviewer for structured scoring, extracts action items, optionally runs GPU experiments for new evidence, revises affected sections, and checks convergence. The loop terminates when the score exceeds a predefined threshold or after a preset maximum of rounds.}
    \label{fig:wf2}
\end{figure*}

\begin{figure*}[t]
    \centering
    \includegraphics[width=0.95\textwidth]{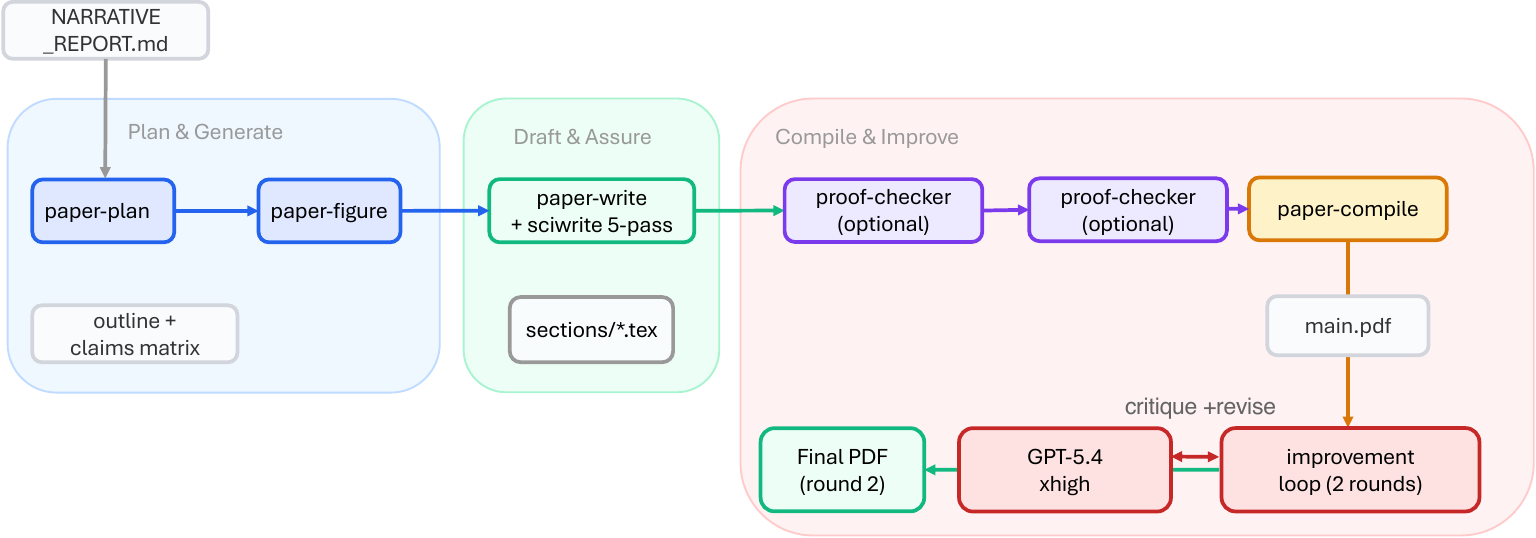}
    \caption{Workflow~3: Paper Writing Pipeline. Three phases: \emph{Plan \& Generate} (outline, figures), \emph{Draft \& Assure} (LaTeX drafting with five-pass editing, optional proof checking, claim auditing), and \emph{Compile \& Improve} (compilation, two rounds of GPT-5.4 xhigh visual review with automatic revision).}
    \label{fig:wf3}
\end{figure*}

Following the harness-engineering taxonomy of \citet{lee2026metaharness}, \aris{} is a \emph{research harness}: a stateful system that orchestrates interactions with LLMs by selecting the context, tools, and feedback presented to them during each stage of a research workflow.
Before describing how the harness is organized internally, we first summarize what it does end-to-end.
Figure~\ref{fig:pipeline} shows the workflow library: five workflows---idea discovery, experiment bridge, auto-review, paper writing, and rebuttal---chained through plain-text artifact contracts and grouped into four research phases (Discovery, Experimentation, Manuscript, Post-Submission).
Figures~\ref{fig:wf2} and~\ref{fig:wf3} zoom into the two assurance-heavy workflows revisited when describing workflow orchestration in \S\ref{sec:realization}: Workflow~2 (Auto Review Loop) and Workflow~3 (Paper Writing).
The architecture, design principles, and adversarial-collaboration mechanism that realize these workflows are described in the remainder of this section; per-skill details follow in \S\ref{sec:realization}.

\begin{figure*}[t]
    \centering
    \includegraphics[width=0.95\textwidth]{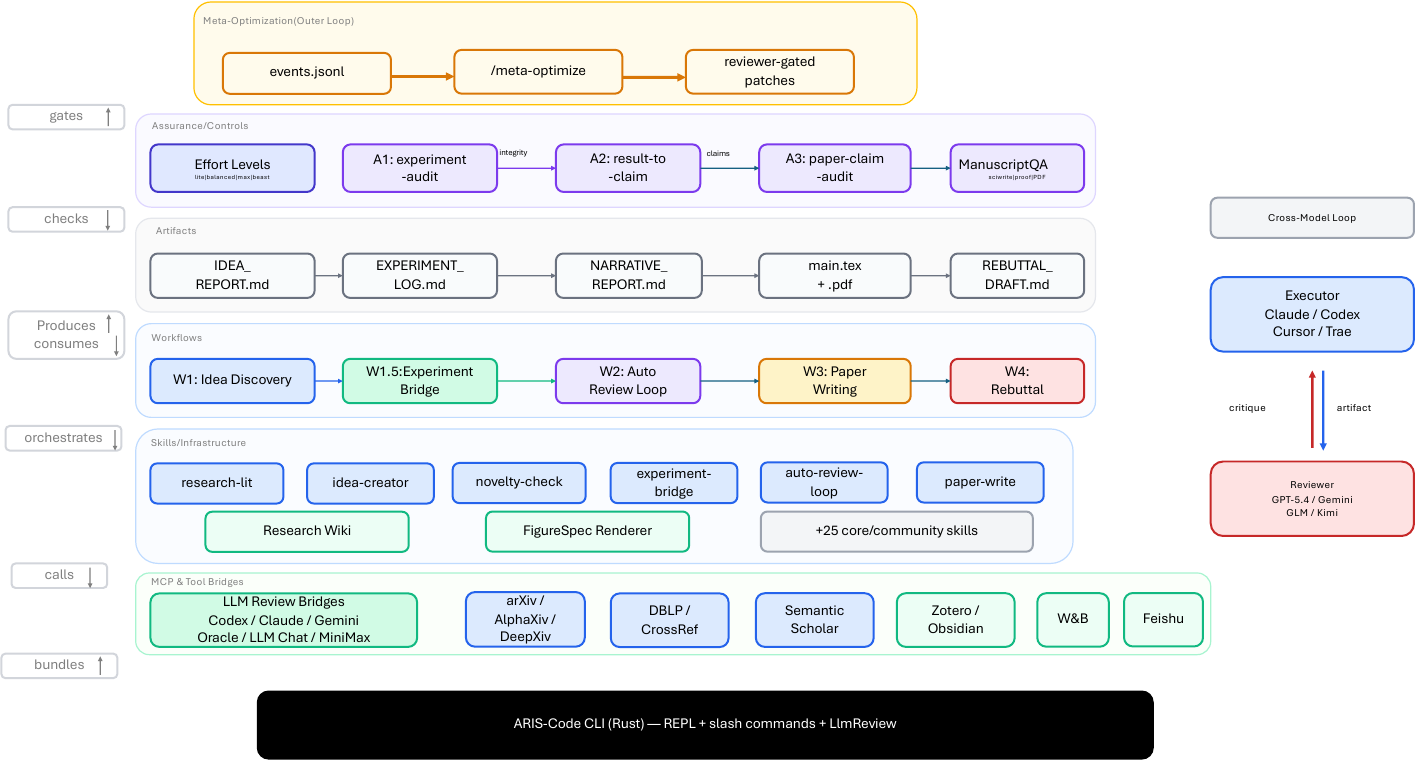}
    \caption{\aris{} system topology. Six component groups interact through labeled relationships (left margin): the \textbf{Meta-Optimization} outer loop gates the \textbf{Assurance} layer, which checks \textbf{Artifacts}; artifacts are produced and consumed by \textbf{Workflows}, which orchestrate \textbf{Skills}; skills call \textbf{MCP \& Tool Bridges} for external model and data access. The executor and reviewer (right) use models from different families. \textbf{ARIS-Code CLI} bundles all components into a standalone binary.}
    \label{fig:arch}
\end{figure*}

Figure~\ref{fig:arch} illustrates the three-layer architecture, and Table~\ref{tab:snapshot} summarizes the implementation described in this report.

\begin{table}[t]
\centering
\caption{Current \aris{} implementation footprint (v0.4, April 2026).}
\label{tab:snapshot}
\small
\begin{tabular}{@{}ll@{}}
\toprule
\textbf{Component} & \textbf{Scope} \\
\midrule
Skills & More than 65 Markdown-defined files \\
Workflows & 5 end-to-end (+ full pipeline command) \\
Model bridges & 6 MCP bridges (Codex, Oracle, Claude, Gemini, MiniMax, llm-chat) \\
Tested executors & 3 (Claude Code, Codex CLI, Cursor); 3 adapted \\
Assurance stack & 3-stage audit cascade + manuscript quality \\
Persistent memory & Per-project research wiki (4 entity types) \\
Effort presets & 4 levels (lite / balanced / max / beast) \\
Dependencies & None for skills; single binary for CLI \\
\bottomrule
\end{tabular}
\end{table}

These layers map to the three bottlenecks identified in \S\ref{sec:intro}: persistent state~(i) is realized by the per-project research wiki and versionable artifact contracts described in \S\ref{sec:wiki}; modular execution~(ii) is realized by self-contained Markdown skill files coordinated through the workflows of Figure~\ref{fig:pipeline}; and independent assurance~(iii) is realized by the assurance layer (\S\ref{sec:assurance}) under the cross-family executor/reviewer pairing detailed below.

\subsection{Design Principles}

The design of \aris{} is guided by five principles.
Principles~(1), (3), and~(5) instantiate bottlenecks~(iii), (ii), and~(i) respectively from \S\ref{sec:intro}; principle~(2) is the implementation choice that makes (ii) ergonomic, and principle~(4) is the engineering constraint that lets these controls survive across executor environments.

\paragraph{(1) Heterogeneous models over single-model self-refinement.}
Single-model self-refinement loops~\citep{madaan2023selfrefine,shinn2023reflexion} have generator and validator that share inductive biases; heterogeneous multi-agent debate has been reported to elicit more diverse critiques than homogeneous configurations~\citep{liang2024divergent,du2024multiagentdebate}.
\aris{} \emph{defaults to} pairing executor and reviewer from different model families and treats this as the recommended configuration.
Here, a \emph{model family} denotes a shared model lineage or provider class (e.g., Claude models form one family; GPT models form another).
The default configuration we ship and document is Claude-family executor with GPT-family reviewer (Codex MCP, Oracle MCP) or vice versa; users can also configure Gemini or MiniMax through dedicated MCP bridges, and GLM, Kimi, or DeepSeek as the reviewer through the generic OpenAI-compatible \texttt{llm-chat} bridge listed in Table~\ref{tab:snapshot}.

\paragraph{(2) Modular skill files over monolithic agents.}
Each research capability is defined primarily by a \skillmd{} file, a plain-text Markdown specification that can be interpreted by multiple LLM-based coding agents, enabling independent development, domain-specific extensions, and component-level updates.

\paragraph{(3) Composability over fixed pipelines.}
Skills can be chained into workflows, with per-invocation parameter overrides and checkpoint-based recovery across sessions.

\paragraph{(4) Portability over vendor lock-in.}
The skill library is distributed as plain-text files and does not depend on a platform-specific runtime; in our current setup, the same \skillmd{} files can be used in Claude Code, Codex CLI, and Cursor with no file-level changes.

\paragraph{(5) Persistent memory over ephemeral context.}
Each project maintains a research wiki that stores papers, ideas, experiment records, and tracked claims across sessions, allowing the system to revisit and refine prior work rather than restarting from a stateless prompt each session~\citep{karpathy2026wiki}.

\subsection{Cross-Model Adversarial Collaboration}
\label{sec:adversarial}

The core mechanism is a \emph{critique-to-action loop}.
The executor first produces an artifact (code, manuscript section, or experiment design).
A reviewer---which the recommended configuration draws from a different model family---then assigns a review score under a predefined rubric and returns structured action items.
The executor addresses those items, after which a convergence check decides whether to run another round or accept the artifact as provisionally satisfactory.
The loop terminates either when the review score exceeds a predefined threshold (default 6/10) and all critical review items have been resolved, or when it reaches a preset maximum number of rounds (default~4).

\begin{figure*}[t]
    \centering
    \includegraphics[width=0.95\textwidth]{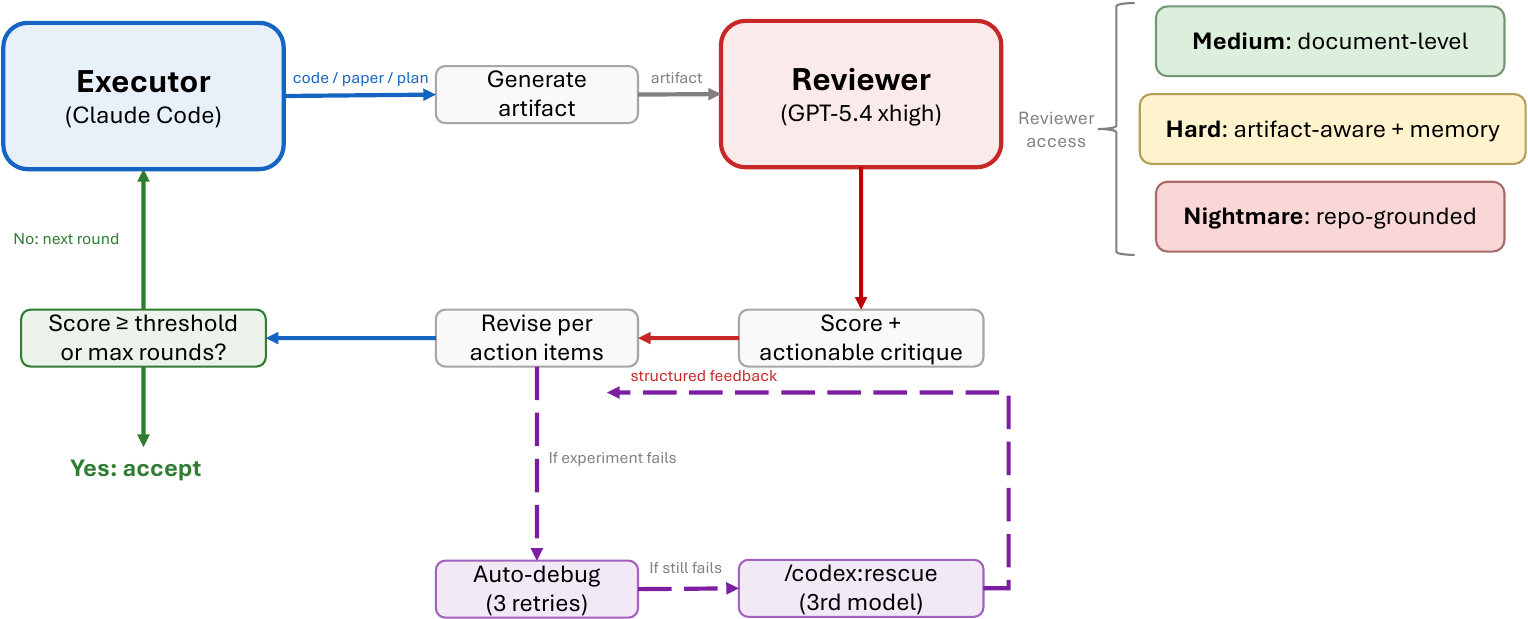}
    \caption{Cross-model adversarial collaboration alternates executor generation with external-model critique, actionable revision requests, and convergence checking. Reviewer access ranges from document-only to repository-level.}
    \label{fig:loop}
\end{figure*}

\paragraph{Reviewer independence.}
The executor supplies file paths and a review objective.
The reviewer then reads the referenced artifacts directly and forms an independent assessment.
If the executor first summarized the artifact, the reviewer would assess the executor's framing rather than the underlying work, thereby increasing the risk of shared errors.
This protocol is specified in a shared protocol document that every skill invoking a review step must follow.

\paragraph{Reviewer access and context policy.}
\aris{} configures reviewers along two orthogonal axes.
The first axis is \emph{access scope}: \emph{document-only} (reviewer reads the manuscript text), \emph{artifact-augmented} (reviewer additionally reads supporting artifacts such as result files), and \emph{repository-level} (reviewer directly inspects the codebase and generated outputs through repository access tools).
The second axis is \emph{context policy}: \emph{fresh} (each review round opens a new thread with no prior context, used to prevent confirmation bias) versus \emph{cross-round} (reviewer retains state across rounds and explicitly verifies whether previously raised issues have been addressed).
Appendix~\ref{app:review} defines each axis in detail and notes which axis settings are required by specific assurance skills.

\paragraph{Automatic debugging and fallback diagnosis.}
When experiments fail, the system assigns the failure to a predefined error class, applies a class-specific remediation, and retries up to a configurable limit (default three attempts).
The executor must attempt at least two distinct remediation strategies before marking a reviewer issue as unresolved.
If both remediation attempts fail, a third, independently configured model can provide an independent diagnosis through a dedicated rescue step.

\section{Cross-Model Assurance Stack}
\label{sec:assurance}

The adversarial collaboration described in \S\ref{sec:adversarial} provides a general critique loop.
It seems perfectly natural that the executor agent only needs to communicate adversarially with the reviewer agent based on the article's content. However, the reality is much more complex. To improve the peer review score as quickly as possible, the executor agent will use various methods to deceive the reviewers during the dialogue. Therefore, we need to set up a strict assurance stack.

This section presents the \emph{assurance stack} that \aris{} adds to the critique loop as its operational response to bottleneck~(iii) of \S\ref{sec:intro} and to the \emph{plausible unsupported success} risk introduced there: a three-stage evidence-to-claim audit cascade for experimental integrity (\S\ref{sec:audit}), a manuscript assurance layer for prose, proof, and presentation quality (\S\ref{sec:manuscript}), and two system-wide controls---effort levels and reviewer routing---that set audit depth and reviewer backend (\S\ref{sec:effort}).

\subsection{Evidence-to-Claim Audit Cascade}
\label{sec:audit}

\begin{figure*}[t]
    \centering
    \includegraphics[width=0.95\textwidth]{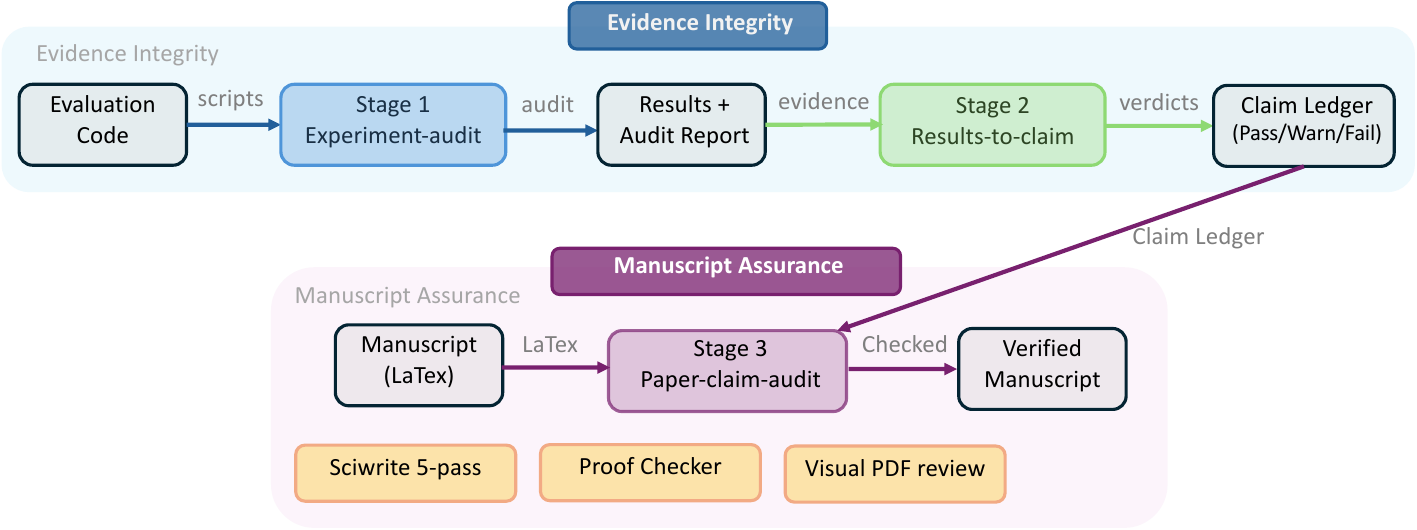}
    \caption{Evidence-to-Claim Audit Cascade. \textbf{Stage~1} (\texttt{experiment-audit}): the reviewer audits evaluation scripts and result files for integrity failure modes. \textbf{Stage~2} (\texttt{result-to-claim}): results are mapped to explicit claim verdicts (supported, partial, invalidated); claims with audit failures are downgraded. \textbf{Stage~3} (\texttt{paper-claim-audit}): a zero-context fresh reviewer compares every quantitative claim in the manuscript against the claim ledger and raw result files. The \textbf{Manuscript Assurance} layer applies four components: a five-pass editing pipeline, proof verification, visual PDF review, and citation-audit (verifying every \texttt{\textbackslash cite} for existence, metadata correctness, and context appropriateness).}
    \label{fig:audit}
\end{figure*}

Community reports and internal debugging revealed that executor agents can produce misleading experimental outputs, including model-derived references, self-normalized metrics, and claims unsupported by output files.
\aris{} addresses these failure modes with a three-stage audit pipeline (Figure~\ref{fig:audit}).
Stage~1 audits evaluation integrity, Stage~2 maps results to explicit claims, and Stage~3 independently verifies manuscript claims against the source and raw evidence using a reviewer that the recommended configuration draws from a model family different from the executor's.

\paragraph{Stage~1: Experiment-integrity audit (\texttt{/experiment-audit}).}
A cross-model reviewer audits the evaluation code and outputs against the following integrity failure modes:
(1)~\emph{model-derived reference labels}---reference targets are synthesized from model outputs rather than obtained from the dataset or another declared source;
(2)~\emph{self-normalized scores}---metrics use denominators derived from the model's own predictions, which can inflate or distort reported performance;
(3)~\emph{phantom results}---claimed numbers that do not match actual output files;
(4)~\emph{dead-code or unused-metric inflation}---evaluation code defines additional metrics or branches that are never executed but are described as part of the analysis;
(5)~\emph{scope inflation}---claims generalize beyond the tested datasets, seeds, or experimental settings.
The audit produces a structured report (\texttt{EXPERIMENT\_AUDIT.md}) and a machine-readable JSON summary.
The audit is advisory at the workflow level: it does not halt execution, but downstream stages propagate warning or failure statuses into later claim judgments.

\paragraph{Stage~2: Result-to-claim mapping (\texttt{/result-to-claim}).}
Each candidate experimental claim is evaluated against the available evidence and assigned one of three verdicts: \emph{supported}, \emph{partially supported}, or \emph{invalidated}.
If a Stage~1 audit report is available, its \texttt{integrity\_status} is propagated to each claim record; claims with \texttt{fail} cannot be marked fully supported until the integrity issue is resolved.
The output is a \emph{claim ledger} that maps each experimental claim to the evidence that supports, qualifies, or contradicts it.

\paragraph{Stage~3: Paper-claim audit (\texttt{/paper-claim-audit}).}
A fresh zero-context reviewer---implemented as a new Codex thread with no prior conversation history---reads the manuscript \LaTeX{} source together with raw result and configuration files, then cross-checks the paper's quantitative claims.
This fresh-thread design reduces the risk that prior executor context or accumulated reviewer expectations bias the audit.
Representative checks include numerical mismatches, best-seed cherry-picking, configuration mismatches between the manuscript and experiment files, aggregation or delta-arithmetic errors, and scope overclaim.
Each claim receives a structured audit status such as \texttt{exact\_match}, \texttt{rounding\_ok}, \texttt{number\_mismatch}, \texttt{config\_mismatch}, or \texttt{missing\_evidence}.

Conceptually, the stages move from code-level integrity, to evidence-to-claim interpretation, to manuscript-level reporting fidelity.
Each stage can be invoked independently.
In the full research pipeline, Stage~1 runs after experiments, Stage~2 assembles claim records from results, and Stage~3 is used during paper writing and final manuscript review.

\subsection{Manuscript Assurance}
\label{sec:manuscript}

Beyond evidence integrity, \aris{} adds four mechanisms for manuscript assurance.

\paragraph{Five-pass scientific-editing pipeline.}
Inspired by the principles of scientific writing pedagogy~\citep{sainani2019writing}, the \texttt{/paper-write} skill applies five automated editing passes after initial drafting:
(1)~\emph{Clutter removal}: remove filler phrases, redundant words, and unnecessary hedging;
(2)~\emph{Active voice}: convert passive constructions to active where appropriate;
(3)~\emph{Sentence structure}: improve topic positioning and local coherence without forcing a single sentence template;
(4)~\emph{Terminology consistency}: if the Methods section introduces a term such as ``validation split,'' later sections should use the same term rather than an informal variant---extract domain-specific key terms and verify consistent usage across sections;
(5)~\emph{Numerical consistency}: cross-check repeated numerical statements against the corresponding table, figure, or cited result file.

\paragraph{Proof verification (\texttt{/proof-checker}).}
For theory-heavy papers, the proof-checker uses a 20-category issue taxonomy together with a two-axis severity scheme that separates proof status (e.g., invalid, unjustified, unclear) from impact (global, local, cosmetic).
The checker verifies theorem applications against side-condition checklists and runs a counterexample red-team pass on key lemmas and major guarantees.
The output is a \emph{proof-obligation ledger} that records the verification status of each theorem, lemma, and derived obligation.

\paragraph{Visual PDF review.}
The \texttt{/auto-paper-improvement-loop} sends \emph{both} the \LaTeX{} source and the compiled PDF to the reviewer.
The reviewer assesses substantive content from the source and visual presentation from the PDF: figure readability, caption--figure alignment, layout quality (orphaned headers, misplaced floats), table formatting, and color consistency across all figures.
This dual-input review catches presentation issues that source-only review misses.

\paragraph{Citation audit (\texttt{/citation-audit}).}
The fourth manuscript-assurance component verifies every \texttt{\textbackslash cite} in the paper along three independent axes:
(i)~\emph{existence}---the cited paper resolves at the claimed arXiv ID, DOI, or venue;
(ii)~\emph{metadata correctness}---author names, year, venue, and title match canonical sources (DBLP, arXiv, ACL Anthology, Nature, OpenReview);
(iii)~\emph{context appropriateness}---the cited paper actually establishes the claim it is being used to support.
The third axis is the most diagnostic: a real paper used to support a wrong claim is a credibility failure that metadata-only checks miss.
Verification uses fresh cross-family reviewers with web access; verdicts are recorded in a per-entry ledger and surfaced as KEEP/FIX/REPLACE/REMOVE recommendations for human approval before submission.

\subsection{Effort Levels and Reviewer Routing}
\label{sec:effort}

\paragraph{Effort levels.}
\aris{} exposes four effort presets that scale breadth-, depth-, and iteration-related settings while leaving core review invariants unchanged:
\texttt{lite} (${\approx}\,0.4\times$) reduces the number of papers surveyed, ideas generated, and review rounds for quick exploration;
\texttt{balanced} ($1\times$, default) provides standard behavior;
\texttt{max} (${\approx}\,2.5\times$) increases search depth, review thoroughness, and experiment repetitions;
\texttt{beast} (${\approx}\,5$--$8\times$) pushes breadth- and iteration-related settings toward their upper bounds.
Users can override the default with an inline directive such as \texttt{effort: max}.
A key invariant is that Codex-based reviewer calls use \texttt{xhigh} reasoning effort regardless of the overall effort preset, so effort scaling changes coverage and iteration counts rather than the reviewer's reasoning budget.

\paragraph{Reviewer routing.}
In the current implementation, review requests route to GPT-5.4 via the Codex MCP bridge.
For especially high-stakes reviews, users can explicitly route supported skills to GPT-5.4 Pro via the Oracle MCP bridge with an inline directive such as \texttt{reviewer: oracle-pro}.
In the current implementation, Oracle routing is enabled for a subset of reviewer-invoking skills.
Alternative reviewer backends can also be connected through the \texttt{llm-chat} bridge, subject to the same reviewer-independence protocol and the recommendation that reviewer and executor come from different model families (\S\ref{sec:overview}).

\section{Implementation: Skills, Workflows, and Tools}
\label{sec:realization}

The assurance layer is covered in \S\ref{sec:assurance}.
This section describes the implementation of the execution and orchestration layers: the skills layer that breaks long research trajectories into inspectable, replaceable units---ARIS's answer to the modular-execution bottleneck~(ii) of \S\ref{sec:intro} (\S\ref{sec:skills}); a per-project research wiki that addresses the persistent-state bottleneck~(i) (\S\ref{sec:wiki}); workflow orchestration (\S\ref{sec:workflows}); and supporting tools (\S\ref{sec:tools}); it then discusses a prototype meta-optimization outer loop (\S\ref{sec:metaopt}).

\subsection{Skills Layer}
\label{sec:skills}

The foundation of \aris{} is a library of more than 65 research-oriented skills (Appendix~\ref{app:skills}), each encoded as a single \skillmd{} file.
A \skillmd{} contains a YAML frontmatter (name, description, trigger conditions, allowed tools) followed by a natural-language workflow specification: inputs, outputs, step-by-step procedures, quality gates, and failure-handling instructions.
Skills range from simple utilities such as \texttt{/arxiv}, which retrieves paper metadata, to multi-step workflows such as \texttt{/auto-review-loop}, which iteratively reviews, revises, and, when needed, runs follow-up experiments.

Five shared reference documents provide cross-cutting guidance: \texttt{reviewer-independence.md}, \texttt{experiment-integrity.md}, \texttt{effort-contract.md}, \texttt{citation-discipline.md}, and \texttt{writing-principles.md}.
Any skill can reference these; they codify system-wide invariants without duplicating rules across skill files.

Skills exchange intermediate artifacts through versionable text files and structured Markdown pages.
For example, \texttt{IDEA\_REPORT.md} is produced during idea discovery and consumed by \texttt{experiment-bridge}; \texttt{EXPERIMENT\_LOG.md} is consumed by \texttt{auto-review-loop}; and \texttt{NARRATIVE\_REPORT.md} is consumed by \texttt{paper-writing}.
This design improves auditability, checkpoint-based recovery, and portability across model backends.
Together, single-file skills and plain-text artifact contracts are how \aris{} discharges bottleneck~(ii) of \S\ref{sec:intro}: the long research trajectory is broken into inspectable, replaceable invocations whose inputs and outputs can be reviewed independently rather than hidden inside a single opaque agent transcript.

\subsection{Research Wiki: Persistent Project Memory}
\label{sec:wiki}

\aris{} realizes bottleneck~(i) of \S\ref{sec:intro}---persistent research state across long-running, multi-session workflows---through four layered mechanisms: (1)~the research wiki described in this subsection, which records papers, ideas, experiments, and claims as a structured knowledge graph; (2)~the plain-text artifact contracts exchanged between skills (\S\ref{sec:skills}), which carry intermediate state across skill invocations; (3)~a \emph{file-system-as-state} design choice (Design Principle~5 of \S\ref{sec:overview}) that places all session state in versionable text files rather than in-memory caches or external databases, so any new session can pick up from the artifacts of a previous one; and (4)~checkpoint-based recovery (Design Principle~3 of \S\ref{sec:overview}), in which any workflow can resume from the saved artifacts of an earlier run.
The wiki is the headline component and is described next; the other three mechanisms are referenced where relevant.

The research wiki provides persistent, cross-session memory through four entity types---papers, ideas, experiments, and claims---stored as structured Markdown pages with canonical node IDs.
Eight typed relationships (extends, contradicts, addresses\_gap, inspired\_by, tested\_by, supports, invalidates, supersedes) form a lightweight knowledge graph.

\begin{figure*}[t]
    \centering
    \includegraphics[width=0.95\textwidth]{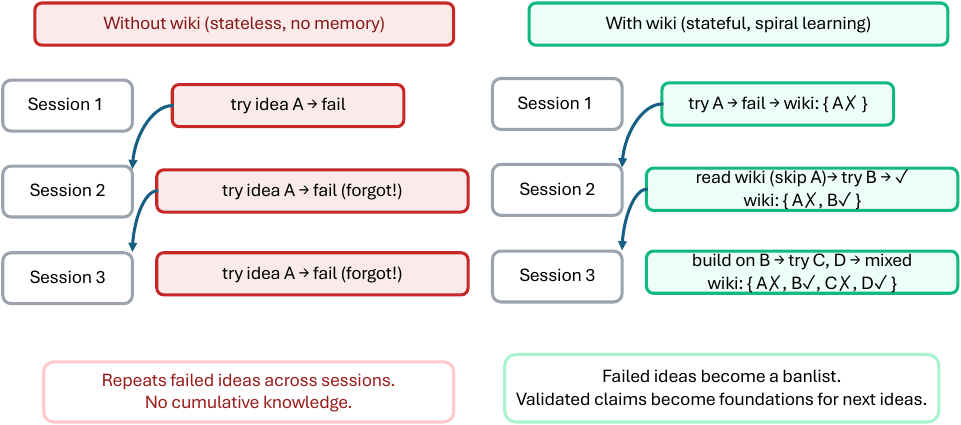}
    \caption{Why the wiki matters. \textbf{Without wiki} (left), each session starts from a blank slate; the same failed idea A can be re-tried indefinitely because the system has no memory of prior outcomes. \textbf{With wiki} (right), Session~1's failure is recorded; Session~2's ideation reads the wiki, skips A, and tries B successfully; Session~3 builds on B and explores C/D. Failed ideas become a banlist; validated claims become foundations for the next ideation round, converting one-shot research into spiral learning.}
    \label{fig:wiki}
\end{figure*}

Three skills integrate with the wiki:
\texttt{/research-lit} ingests discovered papers as structured pages;
\texttt{/idea-creator} reads a compressed \texttt{query\_pack.md} summary (capped at 8{,}000 characters) before ideation, using listed gaps as search seeds and previously rejected ideas to avoid revisiting unpromising directions;
\texttt{/result-to-claim} updates claim status after each experiment.
The key design choice is to retain rejected ideas: without persistent memory, an ideation pipeline can re-propose the same dead-end direction across sessions; with the wiki, the same direction is recognized as previously explored and the search moves on (Figure~\ref{fig:wiki}).

\subsection{Workflow Orchestration}
\label{sec:workflows}

Five workflows chain skills into end-to-end pipelines.
The overall composition is shown earlier in Figure~\ref{fig:pipeline} (\S\ref{sec:overview}); Table~\ref{tab:workflows} lists inputs, outputs, and key skills; full appendix figures for all five workflows are also in Appendix~\ref{app:workflows}.

\begin{table*}[t]
\centering
\caption{\aris{} workflow library. Each workflow chains reusable skills through plain-text artifact contracts. For the idea discovery, the research taste is important for the idea quality, and we recommend the idea taste models provided by \citet{tong2026aitaste}. For experiments, users seeking SoTA results may also find AutoSoTA~\citep{li2026autosota} helpful.}
\label{tab:workflows}
\footnotesize
\begin{tabular}{@{}p{2.6cm}p{2.3cm}p{2.6cm}p{5.5cm}@{}}
\toprule
\textbf{Workflow} & \textbf{Input} & \textbf{Output} & \textbf{Key Skills} \\
\midrule
1. Idea Discovery & Research direction & Ranked idea report & \texttt{research-lit}, \texttt{idea-creator}, \texttt{novelty-check}, \texttt{experiment-plan} \\
1.5. Experiment Bridge & Experiment plan & Running code + results & \texttt{experiment-bridge}, \texttt{run-experiment}, \texttt{monitor-experiment} \\
2. Auto Review Loop & Draft + results & Improved paper & \texttt{auto-review-loop}, \texttt{research-review}, \texttt{analyze-results} \\
3. Paper Writing & Narrative report & Compiled PDF & \texttt{paper-plan}, \texttt{paper-figure}, \texttt{paper-write}, \texttt{proof-checker}, \texttt{paper-claim-audit}, \texttt{paper-compile}, \texttt{auto-paper-improvement-loop} \\
4. Rebuttal & Paper + reviews & Paste-ready rebuttal & \texttt{rebuttal} (7-phase pipeline with 3 safety gates) \\
\bottomrule
\end{tabular}
\end{table*}

\paragraph{Auto-review loop (Workflow~2).}
In each round (Figure~\ref{fig:wf2}, \S\ref{sec:overview}), the draft is sent to a reviewer model from a different family for structured scoring; the system extracts actionable items, runs follow-up experiments when new evidence is requested and execution is permitted, revises affected sections, and resubmits the manuscript for review.
The loop runs for up to four rounds or until the reviewer score exceeds a configurable threshold.
One documented overnight run is described in \S\ref{sec:evidence}.

\paragraph{Paper writing pipeline (Workflow~3).}
This workflow (Figure~\ref{fig:wf3}, \S\ref{sec:overview}) incorporates the assurance components described in \S\ref{sec:assurance}.
The pipeline currently chains seven core sub-skills, with \texttt{/proof-checker} invoked for theory-heavy papers:
\texttt{/paper-plan} produces a structural outline and claims-evidence matrix;
\texttt{/paper-figure} generates manuscript-ready figures and comparison tables;
\texttt{/paper-write} drafts sections in \LaTeX{} with citation lookup and a five-pass revision routine;
optional \texttt{/proof-checker} audits theory-heavy sections;
\texttt{/paper-claim-audit} performs an independent numerical consistency check;
\texttt{/paper-compile} runs multi-pass compilation and repairs common \LaTeX{} errors;
and \texttt{/auto-paper-improvement-loop} performs two rounds of reviewer-model critique followed by revision.
Users can invoke the full writing stack through \texttt{/research-pipeline}; setting \texttt{auto\_write: true} feeds Workflow~2 outputs directly into Workflow~3.

\subsection{Tooling}
\label{sec:tools}

\paragraph{Model bridges.}
\aris{} currently exposes six MCP bridges for executor and reviewer routing: dedicated bridges for Codex, GPT-5.4 Pro review, Gemini, Claude, MiniMax, and a generic OpenAI-compatible chat bridge.
Additional tool bridges cover citation lookup (DBLP/CrossRef), literature search (Semantic Scholar), reference-library sync (Zotero/Obsidian), experiment tracking (W\&B), and mobile notifications (Feishu).

\paragraph{FigureSpec renderer.}
\aris{} includes \texttt{figure\_renderer.py}, a renderer that converts structured JSON FigureSpec descriptions into SVG figures.
The renderer handles shape-aware edge clipping (for rectangular, circular, elliptical, and diamond nodes), self-loops, curved edges, multi-line labels with CJK text width estimation, and comprehensive input validation.
FigureSpec is designed so that LLM agents can generate the JSON programmatically; under a fixed renderer version and font configuration, the same FigureSpec yields the same SVG output.
All architecture and workflow diagrams in this report were generated with this pipeline.

\paragraph{ARIS-Code CLI.}
Beyond skill-based integration into existing IDEs, \ariscode{} is a standalone Rust-based CLI built on \texttt{claw-code}~\citep{ultraworkers2025} that bundles all skills as slash commands.
It ships as a single binary with an interactive REPL, a setup wizard, five LLM providers, and a native \texttt{LlmReview} tool for cross-model critique (Appendix~\ref{app:ariscode}).

\subsection{Meta-Optimization}
\label{sec:metaopt}

Workflows~1--4 optimize research artifacts using a fixed harness.
Meta-optimization targets the harness itself: the skill prompts, default parameters, and convergence rules~\citep{lee2026metaharness}.

\aris{} implements a prototype outer loop in three components:
(1)~\emph{Passive event logging}: in the current prototype, Claude Code hooks record structured events to \texttt{.aris/meta/events.jsonl} during normal usage, including timestamps, tool names, success or failure, and parameter overrides, without requiring manual logging.
(2)~\emph{Pattern analysis}: the \texttt{/meta-optimize} skill analyzes usage statistics---which parameters users override most (suggesting suboptimal defaults), which tools fail repeatedly, where review scores plateau---and proposes targeted patches to the relevant \skillmd{} files.
(3)~\emph{Reviewer-gated application}: each proposed patch is reviewed by GPT-5.4 xhigh; only proposals scoring at least 7/10 are surfaced to the user as recommended candidates. The user makes the final decision; \aris{} never auto-applies harness changes.

\section{Deployment Evidence and Limitations}
\label{sec:evidence}

We summarize deployment footprint and limitations together.
All reported outcomes are \emph{observational}; they cannot be causally attributed to \aris{} alone.

\subsection{Ecosystem and Adoption}

Table~\ref{tab:ecosystem} summarizes the current deployment footprint.
At the time of writing, the skill library had grown from 21 core skills at initial release to more than 65 skills spanning robotics, hardware design, communications, mathematical proof, grant writing, and presentation generation.
At the time of writing, three additional executor environments are documented through community-maintained adaptation guides hosted in external repositories.

\begin{table}[t]
\centering
\caption{Deployment footprint as of April 2026.}
\label{tab:ecosystem}
\small
\begin{tabular}{@{}ll@{}}
\toprule
\textbf{Dimension} & \textbf{Current Status} \\
\midrule
Executor platforms & 3 tested + 3 adapted (6 total) \\
Reviewer models & 6+ (GPT, Gemini, GLM, MiniMax, Kimi, DeepSeek) \\
GPU backends & 4 (local, SSH, Vast.ai, Modal) \\
Venue templates & 9 families \\
Free-tier API access & ModelScope (no paid API keys required) \\
Community contributions & 30+ contributed skills across robotics, hardware, communications, math \\
\bottomrule
\end{tabular}
\end{table}

To illustrate the auto-review loop's operational dynamics under realistic conditions, we documented one overnight run end-to-end.
Over approximately eight hours, the system completed four review--revise rounds, increased an internal reviewer score from 5.0 to 7.5/10, launched more than 20 GPU experiments, and removed claims that were not supported by the available evidence.
This is a single trajectory on one paper; we do not generalize from it.

This run should be read as evidence that the harness can operationalize claim pruning and review-driven revision in one realistic trajectory, not as causal evidence that cross-family review is superior to same-family review or that two cross-family reviewers are an optimal committee size.
The bandit and game-theoretic framing in \S\ref{sec:intro} is used as a design analogy that motivates the two-role pattern; isolating its effect from researcher expertise, model choice, and task difficulty requires the controlled benchmark protocol described in Appendix~\ref{app:benchmark} as future work.

\subsection{Limitations and Responsible Use}

\paragraph{No guarantee of correctness.}
\aris{} cannot guarantee that any output is correct, novel, or scientifically sound.
LLM outputs can include factual hallucinations and methodological gaps; cross-model review reduces some failure modes without eliminating them.
Citation grounding via DBLP and CrossRef reduces but does not eliminate bibliography fabrication; Section~\ref{sec:realization} describes the lookup procedure used in our paper-writing workflow.

\paragraph{Audit limitations.}
The three-stage audit cascade can catch common integrity failures, but it cannot detect every error, inconsistency, or fabrication.
It is an advisory safety net, not a formal verification system.

\paragraph{Reviewer bias amplification.}
The review loop can amplify reviewer biases: if the reviewer consistently demands a particular methodology, the loop may overfit to the reviewer model's preferences rather than improve broader scientific quality.
Over-iteration past diminishing returns can degrade paper quality.

\paragraph{Human responsibility.}
\aris{} automates execution and review loops; humans provide research direction, validate evidence, and make final submission decisions.
Configurable checkpoints (e.g., \texttt{human checkpoint: true}) can be used to require human approval at each workflow step.

\paragraph{Security.}
Repository-level review may send source code to external LLM APIs, raising confidentiality concerns.
Users should not enable repository-level review on repositories containing sensitive code or secrets unless an approved local-only review path is available.
Local-only reviewer routing is planned but not yet implemented.

\paragraph{Self-referential disclosure.}
\aris{} assisted with drafting and review of this technical report, but the authors manually reviewed, edited, and accepted responsibility for all final content.

\section{Related Work}
\label{sec:related}

\paragraph{Autonomous research systems.}
Prior autonomous research systems differ in scope.
The AI Scientist~\citep{lu2024aiscientist} and AI Scientist-v2~\citep{yamada2025aiscientistv2} pursue end-to-end idea-to-paper automation; AI co-scientist~\citep{gottweis2025coscientist} emphasizes hypothesis generation; Agent Laboratory~\citep{schmidgall2025agentrxiv} introduces human-in-the-loop checkpoints; and data-to-paper~\citep{ifargan2024data2paper} targets annotated-data-to-paper workflows with human oversight, programmatic back-tracing, and human-verifiable, information-traceable manuscripts.
These systems differ in how much research state they retain across sessions; some recent systems provide run-level checkpoints or shared research repositories for cumulative progress, such as AgentRxiv~\citep{schmidgall2025agentrxiv}. However, few expose a per-project, structured research memory that jointly records literature notes, ideas, experiments, negative outcomes, and claim status for reuse across sessions.
In contrast, \aris{} defaults to cross-family executor-reviewer separation, ships reusable Markdown skill specifications, maintains a per-project research wiki for persistent cross-session memory of papers, ideas, experiments, and tracked claims (\S\ref{sec:wiki}), and targets portability across multiple executor platforms with limited platform-specific logic.
Recent critical analyses of autonomous research systems~\citep{luo2025aiscientistpitfalls} identify integrity failure modes such as inappropriate benchmark selection, data leakage, metric misuse, and post-hoc selection bias, motivating the explicit assurance machinery we describe in \S\ref{sec:assurance}. Very recently, more interesting auto-research systems, for example, AutoResearchClaw \citep{liu2026autoresearchclaw} and EvoScientist \citep{evoscientist2026} have been built \footnote{Readers can find a broader catalog of auto-research systems at \url{https://cadslab.github.io/Pantheon/}.}.

\paragraph{Self-refinement and multi-agent debate.}
Self-Refine~\citep{madaan2023selfrefine} and Reflexion~\citep{shinn2023reflexion} demonstrate iterative self-feedback and verbal reflection. 
Multi-agent debate~\citep{du2024multiagentdebate} has been reported to improve reasoning in some settings, while divergent-debate work~\citep{liang2024divergent} highlights both the value of forcing alternative arguments and the complications introduced when heterogeneous LLMs participate in judging or debate.
\aris{} draws on these ideas by embedding cross-model review loops throughout the research workflow.
The bandit and two-player game-theoretic language we use in \S\ref{sec:intro} should be read as a design analogy rather than a formal regret or equilibrium result: same-model self-review resembles repeated evaluation under correlated noise, whereas an external reviewer introduces an adversarial role that searches for failure modes the executor did not anticipate.
\aris{} adopts the minimal two-role version of this idea to break self-review blind spots while avoiding the API cost and coordination overhead of larger reviewer committees.

\paragraph{Automated reviewing.}
ReviewerGPT~\citep{liu2023reviewergpt} and large-scale analyses~\citep{liang2024llmfeedback} suggest that LLMs can assist targeted review tasks and produce feedback overlapping with human reviewers on some dimensions, while remaining unsuitable as complete substitutes for expert peer review.
\aris{} uses external-model review as a development tool---iterative improvement during the writing process---not as a substitute for human peer review.

\paragraph{Harness engineering and agent frameworks.}
Meta-Harness~\citep{lee2026metaharness} formalizes outer-loop search over harness code; \aris{} is a hand-engineered research harness with a prototype outer loop as a step in that direction (\S\ref{sec:metaopt}).
AutoGen~\citep{wu2023autogen}, CAMEL~\citep{li2023camel}, OpenHands~\citep{wang2025openhands}, SWE-agent~\citep{yang2024sweagent}, MetaGPT~\citep{hong2024metagpt}, and ChatDev~\citep{qian2024chatdev} are general-purpose agent or software-engineering frameworks.
By contrast, \aris{} focuses on research-specific workflows, domain-aware skill definitions, and reviewer-executor separation across model families.
Table~\ref{tab:comparison} provides a structured comparison.

\begin{table*}[t]
\centering
\caption{Feature comparison. Each column is operationally defined in the caption below; entries reflect features explicitly documented in the cited papers/repos as of our review (April 2026), not author judgment about overall system quality. 
\emph{partial} denotes documented support for a narrower, non-default, or non-identical variant of the feature that does not satisfy the full operational definitions used here.
$\dagger$: tested on 3 platforms (Claude Code, Codex CLI, Cursor) with documented adaptation guides for 3 more.
$\ddagger$: data-driven end-to-end workflow from annotated data to manuscript, rather than open-ended idea-to-paper research.
\textbf{Cross-family policy}: whether the system enforces, defaults to, optionally supports, or does not address cross-family executor/reviewer separation. Entries: \emph{required} (system refuses same-family configurations), \emph{default} (recommended and shipped configuration is cross-family, but not enforced), \emph{optional} (supported but not the default), \emph{none} (no notion of family separation).
\textbf{Adversarial review}: explicit reviewer-vs-executor critique loop with revision.
\textbf{Composable skills}: workflows assembled from independently invocable, single-file skill specifications.
\textbf{E2E research workflows}: covers idea $\to$ experiment $\to$ paper end-to-end.
\textbf{Assurance stack}: explicit, documented integrity/audit mechanisms beyond a single review pass. For this column, \emph{partial} includes narrower provenance, traceability, automated checking, or human-verifiability mechanisms that do not constitute a full assurance stack.
\textbf{Cross-platform portability}: skills usable across multiple host environments without re-implementation.}
\label{tab:comparison}
\footnotesize
\resizebox{\textwidth}{!}{%
\begin{tabular}{@{}lcccccc@{}}
\toprule
\textbf{System} & \textbf{Cross-family} & \textbf{Adversarial} & \textbf{Composable} & \textbf{E2E Research} & \textbf{Assurance} & \textbf{Cross-platform} \\
& \textbf{policy} & \textbf{review} & \textbf{skills} & \textbf{workflows} & \textbf{stack} & \textbf{portability} \\
\midrule
AI Scientist~\citep{lu2024aiscientist} & none & partial & \texttimes & \checkmark & partial & \texttimes \\
AI Scientist-v2~\citep{yamada2025aiscientistv2} & none & partial & \texttimes & \checkmark & partial & \texttimes \\
Agent Laboratory~\citep{schmidgall2025agentrxiv} & none & \texttimes & \texttimes & \checkmark & \texttimes & \texttimes \\
data-to-paper~\citep{ifargan2024data2paper} & none & partial & \texttimes & \checkmark$^{\ddagger}$ & partial & \texttimes  \\
AutoGen~\citep{wu2023autogen} & none & \texttimes & partial & \texttimes & \texttimes & \texttimes \\
MetaGPT~\citep{hong2024metagpt} & none & partial & partial & \texttimes & \texttimes & \texttimes \\
OpenHands~\citep{wang2025openhands,openhands_skills} & none & \texttimes & partial & \texttimes & \texttimes & \texttimes \\
\textbf{\aris{} (ours)} & default & \checkmark & \checkmark & \checkmark & \checkmark & \checkmark$^\dagger$ \\
\bottomrule
\end{tabular}%
}
\end{table*}
\section{Conclusion}
\label{sec:conclusion}

This report presented \aris{} as a research harness built around a conservative assumption: long-horizon research performed by a single agent is unreliable by default, and the relevant failure mode is not visible breakdown but \emph{plausible unsupported success}, where claims outrun evidence and later readers silently inherit the executor's framing.
\aris{} responds by decomposing the workflow into the three bottlenecks framed in \S\ref{sec:intro}---persistent research state, modular execution, and independent assurance---and by adopting a two-role cross-family reviewer-executor pattern as the practical minimum for breaking self-review blind spots.
These three bottlenecks map to three layers: an execution layer of reusable Markdown-defined skills and a persistent research wiki, an orchestration layer for configurable workflow control and reviewer routing, and an assurance layer for evidence-to-claim auditing and manuscript checks.
A prototype meta-optimization loop provides an initial mechanism for improving skill prompts, defaults, and convergence rules over time.

The main limitations are the absence of controlled evaluation and the reliance on observational deployment evidence.
Future work includes compute-matched comparisons to estimate the contribution of cross-model heterogeneity (Appendix~\ref{app:benchmark}), local reviewer models for confidential settings, and user studies of researcher productivity.

As a more speculative adjacent direction, the cross-model accountability primitives developed in \aris{}---reviewer independence, evidence-to-claim audit, and provenance-aware claim ledgers---are not specific to manuscripts.
A natural adaptation is to insert them between any model output and any downstream training-data retention or reward signal, complementing recent self-improvement approaches~\citep{bai2022constitutional,lee2024rlaif,yuan2024selfrewarding,yu2025rlaifv} with an explicit oversight layer.
Two known concerns motivate the hypothesis: LLM judges can exhibit systematic biases~\citep{zheng2023judging}, and recursive training on model-generated data can degrade quality across iterations~\citep{shumailov2024curse}; cross-family reviewer separation is a candidate mechanism for reducing judge-model coupling, but its downstream effect on long-horizon self-improvement remains an open empirical question.
This is a testable future-work hypothesis, not a claim made in this report.

Code and documentation are available at \url{https://github.com/wanshuiyin/Auto-claude-code-research-in-sleep}.

\bibliography{references}
\bibliographystyle{iclr2026_conference}

\newpage
\appendix

\section{Workflow Internals}
\label{app:workflows}

\begin{figure*}[h]
    \centering
    \includegraphics[width=0.95\textwidth]{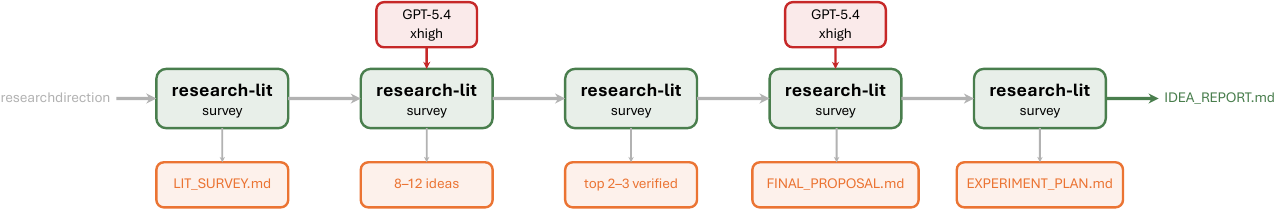}
    \caption{Workflow~1: Idea Discovery. The pipeline surveys literature, brainstorms ideas via cross-model generation, verifies novelty, and refines the top proposal through iterative GPT-5.4 review.}
    \label{fig:appwf1}
\end{figure*}

\begin{figure*}[h]
    \centering
    \includegraphics[width=0.95\textwidth]{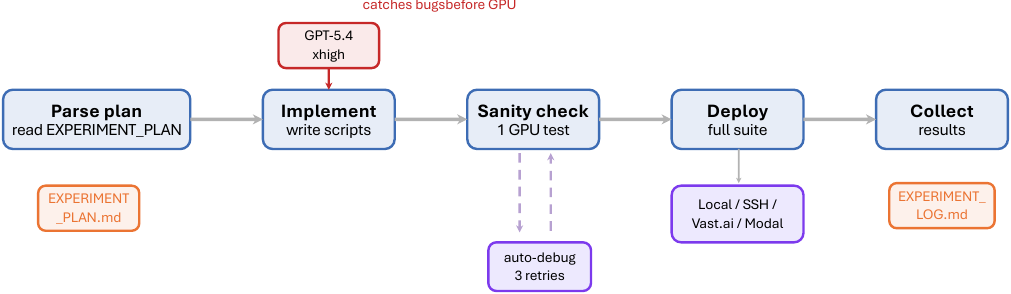}
    \caption{Workflow~1.5: Experiment Bridge. Scripts are implemented, reviewed for code correctness, sanity-checked on one GPU, then deployed to the full backend.}
    \label{fig:appwf15}
\end{figure*}

\begin{figure*}[h]
    \centering
    \includegraphics[width=0.95\textwidth]{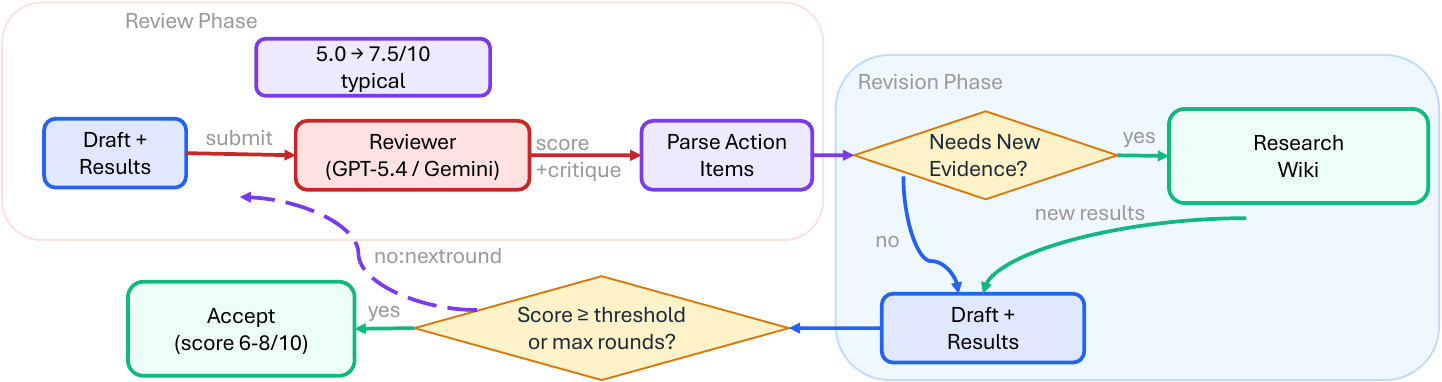}
    \caption{Workflow~2: Auto Review Loop. The reviewer scores the manuscript, the executor implements fixes and runs requested experiments, and the cycle repeats.}
    \label{fig:appwf2}
\end{figure*}

\begin{figure*}[h]
    \centering
    \includegraphics[width=0.95\textwidth]{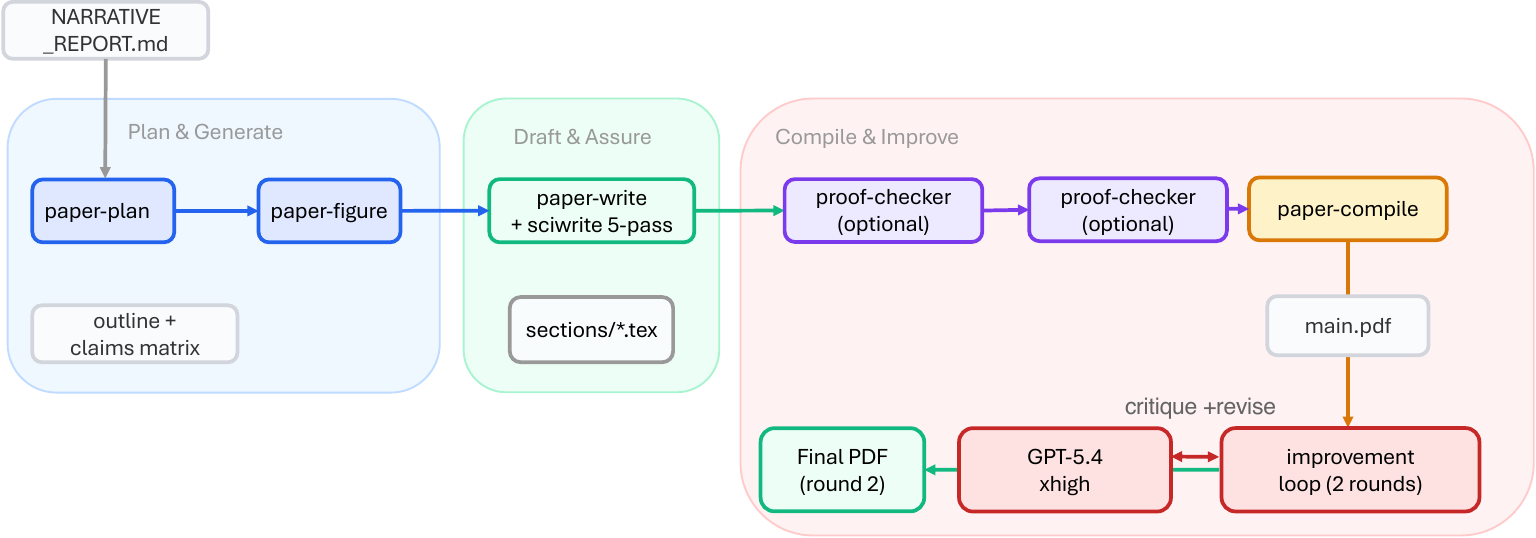}
    \caption{Workflow~3: Paper Writing Pipeline. Seven core sub-skills (plus optional proof checking) chain from outline through figure generation, \LaTeX{} drafting, claim auditing, compilation, and review.}
    \label{fig:appwf3}
\end{figure*}

\begin{figure*}[h]
    \centering
    \includegraphics[width=0.95\textwidth]{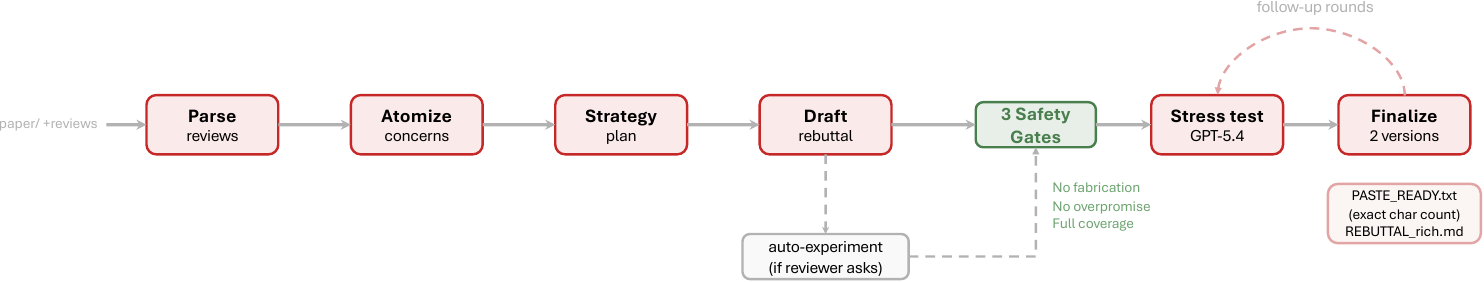}
    \caption{Workflow~4: Rebuttal. Seven phases from parsing reviews through stress-testing, with three safety gates.}
    \label{fig:appwf4}
\end{figure*}

Figures~\ref{fig:appwf1}--\ref{fig:appwf4} show the internal structure of each workflow.

\section{Skill Inventory}
\label{app:skills}

Table~\ref{tab:fullinventory} lists core framework skills in the current release.

\begin{table*}[h]
\centering
\caption{Core \aris{} skill inventory (v0.4, April 2026). Community-contributed skills (30+) are omitted for brevity.}
\label{tab:fullinventory}
\footnotesize
\resizebox{\textwidth}{!}{%
\begin{tabular}{@{}llll@{}}
\toprule
\textbf{Skill} & \textbf{Category} & \textbf{Reviewer?} & \textbf{Key Function} \\
\midrule
\texttt{research-lit} & Literature & No & Multi-source literature survey \\
\texttt{arxiv} / \texttt{alphaxiv} & Literature & No & Paper metadata / LLM-optimized summary \\
\texttt{deepxiv} & Literature & No & Progressive deep literature search \\
\texttt{novelty-check} & Literature & Yes & Novelty verification against existing work \\
\texttt{idea-creator} & Ideation & Yes & Brainstorm and rank research ideas \\
\texttt{idea-discovery} & Ideation & Yes & Full idea discovery pipeline \\
\texttt{experiment-bridge} & Experiment & Yes & Plan to running code \\
\texttt{run-experiment} & Experiment & No & GPU deployment (local/SSH/Vast.ai/Modal) \\
\texttt{monitor-experiment} & Experiment & No & Experiment monitoring and result collection \\
\texttt{check-gpu} & Experiment & No & GPU status and process summary \\
\texttt{analyze-results} & Analysis & No & Statistical analysis and comparison tables \\
\texttt{ablation-planner} & Analysis & No & Reviewer-perspective ablation design \\
\texttt{experiment-audit} & Integrity & Yes & Evaluation code integrity verification \\
\texttt{result-to-claim} & Integrity & No & Result-to-claim verdict conversion \\
\texttt{paper-claim-audit} & Integrity & Yes & Zero-context manuscript number audit \\
\texttt{proof-checker} & Integrity & No & 20-category theorem verification \\
\texttt{auto-review-loop} & Review & Yes & Multi-round autonomous review \\
\texttt{research-review} & Review & Yes & GPT-5.4 xhigh deep critique \\
\texttt{paper-plan} & Writing & Yes & Structural outline + claims matrix \\
\texttt{paper-write} & Writing & Yes & Section-by-section \LaTeX{} + sciwrite 5-pass \\
\texttt{paper-figure} & Writing & No & Publication-quality data plots \\
\texttt{paper-compile} & Writing & No & Multi-pass compilation with auto-repair \\
\texttt{paper-writing} & Writing & Yes & Full W3 pipeline orchestrator \\
\texttt{auto-paper-improvement-loop} & Writing & Yes & Review + visual PDF + auto-revision \\
\texttt{rebuttal} & Writing & Yes & 7-phase rebuttal with safety gates \\
\texttt{research-wiki} & Memory & No & Persistent knowledge base management \\
\texttt{meta-optimize} & Maintenance & Yes & Outer-loop harness optimization \\
\bottomrule
\end{tabular}%
}
\end{table*}

\section{Reviewer Configuration}
\label{app:review}

\aris{} configures reviewer behavior along two orthogonal axes (Section~\ref{sec:adversarial}). Table~\ref{tab:scope} lists the three access-scope settings; Table~\ref{tab:context} lists the two context-policy settings.

\begin{table}[h]
\centering
\caption{Reviewer access-scope settings (what the reviewer is allowed to read).}
\label{tab:scope}
\small
\begin{tabular}{@{}lp{8cm}@{}}
\toprule
\textbf{Scope} & \textbf{Description} \\
\midrule
Document-only & Reviewer reads only the manuscript text. Default for paper-writing review. \\
Artifact-augmented & Reviewer additionally reads result files, claim ledgers, and intermediate artifacts referenced by the manuscript. \\
Repository-level & Reviewer directly inspects the codebase, evaluation scripts, and generated outputs through host-provided repository access tools (e.g., \texttt{claude-code} Bash, \texttt{codex exec}). Used by \texttt{experiment-audit} and \texttt{paper-claim-audit}. \\
\bottomrule
\end{tabular}
\end{table}

\begin{table}[h]
\centering
\caption{Reviewer context-policy settings (whether the reviewer retains state across rounds).}
\label{tab:context}
\small
\begin{tabular}{@{}lp{8cm}@{}}
\toprule
\textbf{Policy} & \textbf{Description} \\
\midrule
Fresh & Each review round opens a new reviewer thread with no prior context. Used to prevent confirmation bias from previous rounds (\texttt{REVIEWER\_BIAS\_GUARD = true} default). Required for \texttt{paper-claim-audit} and the auto-paper improvement loop. \\
Cross-round & Reviewer retains memory across rounds, can reference previous critiques and verify whether raised issues have been addressed. Used selectively when convergence verification is more important than independence. \\
\bottomrule
\end{tabular}
\end{table}

\section{ARIS-Code Details}
\label{app:ariscode}

\ariscode{} is a standalone Rust-based CLI built on \texttt{claw-code}~\citep{ultraworkers2025}.
Key features: interactive REPL with setup wizard, all skills as slash commands, five LLM providers, native \texttt{LlmReview} tool, three-tier skill priority system (user $>$ Claude Code $>$ bundled), and \texttt{/cost} for token tracking.

\section{Controlled Benchmark Protocol (Future Work)}
\label{app:benchmark}

We outline a benchmark protocol for future controlled evaluation:
\textbf{Task pool}: 12+ paper drafts from publicly available preprints.
\textbf{Conditions} (compute-matched): (A)~single-model self-critique, (B)~same-model two-agent, (C)~cross-model, (D)~cross-model reversed, (E)~same-model for the second model.
\textbf{Metrics}: issue recall, false-positive rate, actionability score, downstream revision quality, cost, latency.
\textbf{Raters}: three independent, blinded. Inter-rater agreement via Krippendorff's $\alpha$.

\end{document}